\documentclass[]{aa}  
\usepackage{graphicx}
\usepackage{txfonts}
\usepackage{hyperref}
\hypersetup{
    colorlinks,
    citecolor=blue,
    filecolor=blue,
    linkcolor=blue,
    urlcolor=blue}

\usepackage{float}
\usepackage{amsmath}
\usepackage{natbib}
\usepackage{silence}
\newcommand{\co}[2]{\(\rm CO~({#1} - {#2})\)}
\newcommand{\ci}[2]{\(\rm [CI]~({#1} - {#2})\)}

\newcommand{\sig}[1]{\({#1} \sigma\)}
\newcommand{\val}[2]{\(\rm {#1}~{#2}\)}
\newcommand{\simval}[2]{\(\sim\)\(\,\)\(\rm {#1}~{#2}\)}
\newcommand{\appval}[2]{\(\approx\)\(\,\)\(\rm {#1}~{#2}\)}

\begin{document} 

    \title{Deep kiloparsec view of the molecular gas in a massive star-forming galaxy at cosmic noon}


    \author{Sebastián Arriagada-Neira\inst{1}
            \and
            Rodrigo Herrera-Camus\inst{1}
            \and
            Vicente Villanueva\inst{1}
            \and
            Natascha M. Förster Schreiber\inst{2}
            \and
            Minju Lee\inst{3,4}
            \and
            Alberto Bolatto\inst{5}
            \and
            Jianhang Chen\inst{2}
            \and
            Reinhard Genzel\inst{2}
            \and
            Daizhong Liu\inst{6}
            \and
            Alvio Renzini\inst{7}
            \and
            Linda J. Tacconi\inst{2}
            \and
            Giulia Tozzi\inst{2}
            \and
            Hannah Übler\inst{8,9}
           }

    \authorrunning{Arriagada-Neira et al.}

    \institute{Departamento de Astronomía, Universidad de Concepción, Barrio Universitario, Concepción, Chile\\
              \email{searriagada2018@udec.cl}
              \and
              Max-Planck-Institut für extraterrestische Physik (MPE), Giessenbachstr., D-85748 Garching, Germany
              \and
              Cosmic Dawn Center (DAWN), Denmark
              \and
              DTU-Space, Technical University of Denmark, Elektrovej 327, DK-2800 Kgs. Lyngby, Denmark
              \and
              Department of Astronomy, University of Maryland, College Park, MD 20742, USA
              \and
              Purple Mountain Observatory, Chinese Academy of Sciences, 10 Yuanhua Road, Nanjing 210023, People’s Republic of China
              \and
              INAF - Osservatorio Astronomico di Padova, Vicolo dell'Osservatorio 5, I-35122 Padova, Italy
              \and
              Kavli Institute for Cosmology, University of Cambridge, Madingley Road, Cambridge CB3 0HA, UK
              \and
              Cavendish Laboratory – Astrophysics Group, University of Cambridge, 19 JJ Thomson Avenue, Cambridge CB3 0HE, UK
              }

    \date{Received XXXX; accepted XXXX}

    \abstract
    {We present deep (\(\sim\)\(\,\)\(\rm {20}~{hr}\)), high-angular resolution Atacama Large Millimeter/submillimeter Array (ALMA) observations of the \(\rm CO ~ (4-3)\) and \(\rm [CI] ~ (1-0)\) transitions, along with the rest-frame \(\rm 630 ~{\rm {\mu}m}\) dust continuum, in BX610—a massive, main-sequence galaxy at the peak epoch of cosmic star formation \((z = 2.21)\). Combined with deep Very Large Telescope (VLT) SINFONI observations of the \(\rm H{\alpha}\) line, we characterize the molecular gas and star formation activity on kiloparsec scales. Our analysis reveals that the excitation of the molecular gas, as traced by the \(L'_{\rm CO~(4-3)} ~ \big/ ~ L'\:\!\!_{\rm [CI]~(1-0)}\) line luminosity ratio, decreases with increasing galactocentric radius. While the line luminosity ratios in the outskirts are similar to those typically found in main-sequence galaxies at \(z \sim 1\), the ratios in the central regions of BX610 are comparable to those observed in local starbursts. There is also a giant extra-nuclear star-forming clump in the southwest of BX610 that exhibits high star formation activity, molecular gas abundance, and molecular gas excitation. Furthermore, the central region of BX610 is rich in molecular gas \((M_{\rm mol} / M_{\rm \star} \approx 1)\); however, at the current level of star formation activity, such molecular gas is expected to be depleted in \(\sim\)\(\,\)\(450~ {\rm Myr}\). This, along with recent evidence for rapid inflow toward the center, suggests that BX610 may be experiencing an evolutionary phase often referred to as wet compaction, which is expected to lead to central gas depletion and subsequent inside-out quenching of star formation activity.}


    \maketitle


\section{Introduction}
    The epoch known as cosmic noon, when the Universe was \simval{2-3}{Gyr} old, represents the peak of the cosmic star formation rate (SFR) density \citep[e.g.,][]{madau+14}, making it a critical period for studying galaxy evolution. Recent galaxy surveys have shown that the increased star formation activity during this time was primarily driven by abundant reservoirs of cold molecular gas, the fuel for star formation \citep[e.g.,][]{Tacconi+18, Walter+20, Tacconi+20}. Notably, the molecular gas fraction in massive star-forming galaxies \((M_{\star} \gtrsim 10^{10.5} ~ {\rm M_{\odot}})\) at \(z \sim 1-3\) was close to unity \((M_{\rm mol} / M_{\star} \approx 1)\), highlighting the important role of molecular gas in sustaining the observed high star formation rates during this epoch.

    A new generation of mm/sub-mm facilities, including the Atacama Large Millimeter/submillimeter Array (ALMA), the NOrthern Extended Millimeter Array (NOEMA), and the Karl G. Jansky Very Large Array (JVLA), has provided the advanced capabilities necessary for detailed studies of molecular gas properties in normal, star-forming galaxies at high redshift. The most widely used molecular gas tracer has been the carbon monoxide (CO) molecule \citep[e.g.,][]{Bolatto+13}. However, in recent years, the lower fine-structure line of atomic carbon \(\rm [CI]\) \((^{3}P_{1}\) \(\to\) \(^{3}P_{0})\) (hereafter \ci{1}{0}) has emerged as an alternative molecular gas tracer \citep[e.g.,][]{Papadopoulos+04, Alaghband-Zadeh+13, Bothwell+17, Michiyama+21}. Both theoretical \citep[e.g.,][]{Offner+14, Glover+15} and observational \citep[e.g.,][]{Ikeda+02, Kulesa+05, Requena-Torres+16, Okada+19} studies have shown that [CI] emission is associated with CO emission, regardless of the environment, making it a potential robust tracer of the bulk molecular gas mass \citep{Papadopoulos+04, Walter+11, Henríquez-Brocal+22}. Even in metal-poor and high cosmic-ray environments where CO could be significantly photodissociated, [CI] emission can persist throughout the molecular gas cloud \citep[e.g.,][]{Papadopoulos+04, Bisbas+15, Bisbas+17, Glover+16}. Furthermore, studies by \cite{Valentino+18, Valentino+20} have identified a tight correlation between [CI] and CO emission lines that is independent of redshift (for \(z \lesssim 3\)) and galaxy type. They also found that the [CI] transition can trace molecular gas mass comparably to low-\textit{J} CO lines and dust continuum emission.

    To date, estimates of molecular gas content in high-\(z\) galaxies have been obtained either from large-scale surveys of star-forming galaxies \citep[e.g.,][]{Genzel+10, Tacconi+13, Daddi+10, Valentino+18, Valentino+20}, or from a limited number of spatially resolved studies focusing on various subset of galaxies, including bright submillimeter galaxies \citep[SMGs; e.g.,][]{Hodge+15, Chen+17}, protocluster galaxies at high redshift \citep[e.g.,][]{Lee+17, Lee+21}, cluster galaxies \citep[e.g.,][]{Ikeda+22}, gravitationally lensed dusty galaxies \citep[e.g.,][]{Spilker+15}, and compact star-forming galaxies \citep[e.g.,][]{Spilker+19}. Despite significant progress, the molecular gas properties of more typical star-forming galaxies at cosmic noon on \(\sim\)kiloparsec scales remain relatively understudied. One area where progress has been made is in understanding the spatially resolved Kennicutt-Schmidt (KS) relation \citep{Kennicutt+98a} between total molecular gas and SFR. Studies, for example, by \cite{Genzel+13} and \cite{Freundlich+13} have demonstrated that both the slope of the KS relation and the molecular gas depletion timescale \((t_{\rm dep} = M_{\rm mol} / {\rm SFR})\) in galaxies at cosmic noon are broadly consistent with those observed in the local Universe \citep[e.g.,][]{Bolatto+17, Sun+23}.

    Motivated by the need to characterize the molecular gas properties on \(\sim\)kiloparsec scales of representative galaxies at cosmic noon, we present the deepest ALMA observation to date of a typical main-sequence galaxy at \(z \sim 2\). The focus of our study, Q2343-BX610 (hereafter BX610), is a main-sequence star-forming galaxy at \(z = 2.2103\) with a stellar mass of \(M_{\star} = 1.55 \times 10^{11} ~ {\rm M_{\odot}}\) \citep{Tacchella+15} and a SFR ranging from \(60\) to \(200 ~ {\rm M_{\odot}~yr^{-1}}\), depending on the tracer or the level of dust obscuration assumed \citep{ForsterSchreiber+09, ForsterSchreiber+14, Brisbin+19}. BX610 has been extensively studied across various wavelengths, including optical/near-infrared \citep{ForsterSchreiber+09, ForsterSchreiber+11A, ForsterSchreiber+11B, ForsterSchreiber+14, ForsterSchreiber+18} and far-infrared \citep{Tacconi+13, Aravena+14, Bolatto+15, Brisbin+19}. \(\rm H{\alpha}\) observations characterize BX610 as a rotating disk with a massive star-forming clump and no evidence of a nearby companion \citep{ForsterSchreiber+11B, ForsterSchreiber+18}. The galaxy rest-frame optical morphology reveals bar- and spiral-like features \citep{ForsterSchreiber+11A}. Additionally, observational evidence suggests that BX610 may host an active galactic nucleus (AGN), but it appears to be relatively weak \citep{ForsterSchreiber+14, newman+14}.

    Previous observational efforts to investigate the physical conditions of the molecular gas in BX610 include JVLA \co{1}{0} observations by \cite{Bolatto+15}, and IRAM Plateau De Bure Interferometer (PdBI) \(\rm CO~(4-3, ~7-6)\) and \(\rm [CI]~(1-0, ~2-1)\) observations by \cite{Brisbin+19}. However, due to depth and/or angular resolution limitations, these studies have provided only a broad overview of BX610 molecular gas properties. In this context, the new deep (\simval{20}{hr} of ALMA time), \(\sim\)kiloparsec scale ALMA observations presented in this work represent a significant improvement in depth and spatial resolution. The kinematic properties of BX610 based on this dataset have already been analyzed by \cite{Genzel+23}, who find that the velocity field of BX610 traced by the \co{4}{3} emission line reveals a well-defined rotating disk, with evidence for a rapid inflow of gas towards the central region. The velocity of the radial gas motion is \simval{95}{km~s^{-1}}, approximately one-third of the rotational velocity.

    This paper is organized as follows. In Section \ref{sec:sec02}, we present the new ALMA observations and the further data reduction, as well as the archive observations used in this work. In Section \ref{sec:sec03}, we explain the methods applied to analyze the data and the equations used to derive the physical quantities. In Section \ref{sec:sec04}, we present our results and discussion, and finally, in Section \ref{sec:sec05}, we summarize the main conclusions. Throughout this paper, we assume a Chabrier initial mass function (IMF; \citealt{Chabrier+03}) and a \(\rm {\Lambda}CDM\) cosmology with \(H_{0} = 70 ~ {\rm km~s^{-1}~Mpc^{-1}}\), \(\Omega_{\rm M} = 0.3\), and \(\Omega_{\Lambda} = 0.7\). For this cosmology, \(1\arcsec\) corresponds to \val{8.259}{kpc} at \(z = 2.2103\).


\section{Data Products} \label{sec:sec02}
    \begin{figure*}
        \centering
        \includegraphics{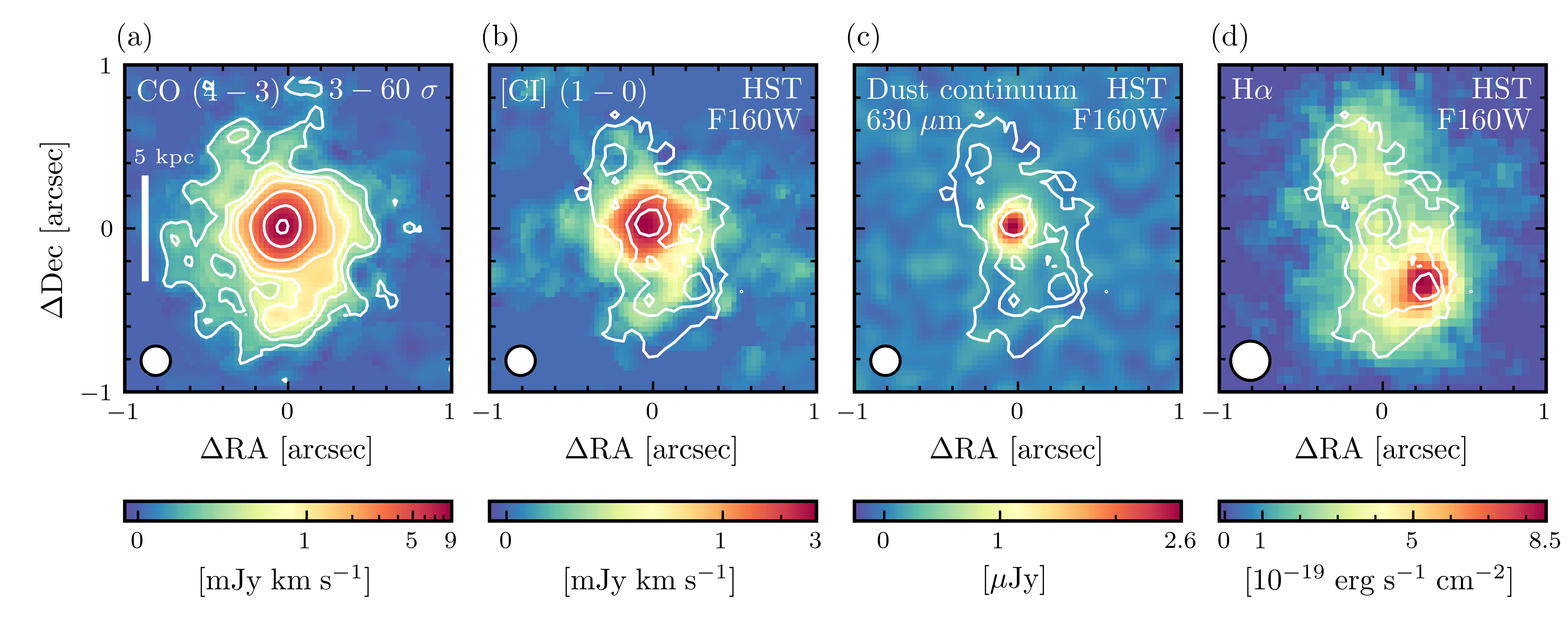}
        \caption{Multi-wavelength imaging of BX610. (a) \co{4}{3} integrated intensity map. The overlaid white contours correspond to the 3, 5, 7, 10, 20, 40, and \sig{60} levels. The filled circle at the bottom left corner indicates the ALMA-synthesized beam \(\rm (\theta = 0.18\arcsec \times 0.18\arcsec)\). (b) \ci{1}{0} integrated intensity map. The overlaid white contours, as in panels (c) and (d), represent the rest-frame optical emission observed by HST WFC3 F160W, corresponding to fractional flux levels relative to the maximum of 0.25, 0.5, and 0.75. (c) Rest-frame \val{630}{\mu m} dust continuum emission. (d) \(\rm H{\alpha}\) integrated intensity map. The filled circle at the bottom left corner indicates the \(\rm SINFONI ~ H{\alpha}\) point-spread function \(\rm (\theta = 0.24\arcsec \times 0.24\arcsec)\)}.
        \label{fig:fig01}
    \end{figure*}
    \begin{table}
        \tiny
        \setlength{\tabcolsep}{4pt}
        \caption[]{Observed and derived properties of BX610.}
        \label{tab:tab01}
        \centering
        \begin{tabular}{lccc}
        \hline
        \hline
        &&&\\ [-1.75ex]
        Parameter & Unit & Value & Reference\\ [0.41ex]
        \hline
        &&&\\ [-1.75ex]
        RA                                 & hh:mm:ss.ss               & 23:46:09.43         & \cite{Erb+06}\\ [0.41ex]
        Dec                                & dd:mm:ss.ss               & 12:49:19.21         & \cite{Erb+06}\\ [0.41ex]
        Redshift                           &                           & 2.2103              & \cite{ForsterSchreiber+09}\\ [0.41ex]
        \(R_{\rm eff, \,\star}\)           & \(\rm kpc\)               & \(4.44 \pm 0.08\)   & \cite{ForsterSchreiber+11A}\\ [0.41ex]
        \(S_{\rm CO~(1-0)} {\Delta}{v}\)   & \(\rm Jy~km~s^{-1}\)      & \(0.105 \pm 0.015\) & \cite{Bolatto+15}\\ [0.41ex]
        FWHM                               & \(\rm km~s^{-1}\)         & \(294 \pm 49 \)     & \cite{Bolatto+15}\\ [0.41ex]
        \(S_{\rm CO~(4-3)} {\Delta}{v}\)   & \(\rm Jy~km~s^{-1}\)      & \(1.55 \pm 0.03\)   & This work\\ [0.41ex]
        FWHM                               & \(\rm km~s^{-1}\)         & \(280 \pm 12\)      & This work\\ [0.41ex]
        \(S_{\rm [CI]~(1-0)} {\Delta}{v}\) & \(\rm Jy~km~s^{-1}\)      & \(0.54 \pm 0.04\)   & This work\\ [0.41ex]
        FWHM                               & \(\rm km~s^{-1}\)         & \(296 \pm 32\)      & This work\\ [0.41ex]
        \(S_{\rm 630 ~ {{\mu}m}}\)         & \(\rm mJy\)               & \(0.18 \pm 0.01\)   & This work\\ [0.41ex]
        \(M_{\star}\)                      & \(\rm 10^{11}~M_{\odot}\) & \(1.55\)            & \cite{Tacchella+15}\\ [0.41ex]
        \(\rm SFR(SED)\)                   & \(\rm M_{\odot}~yr^{-1}\) & \(144 \pm 6\)       & \cite{Brisbin+19}\\ [0.41ex]
        \(M_{\rm mol}{\rm (CO)}\)          & \(\rm 10^{11}~M_{\odot}\) & \(1.2 \pm 0.2\)     & This work\\ [0.41ex]
        \(M_{\rm mol}{\rm ([CI])}\)        & \(\rm 10^{11}~M_{\odot}\) & \(1.3 \pm 0.1\)     & This work\\ [0.41ex]
         \hline
        \end{tabular}
    \end{table}
    To analyze the physical conditions of the ISM in BX610, we use ALMA observations along with archival data from the Very Large Telescope (VLT) and Hubble Space Telescope (HST). The key features of these datasets are summarized in the following subsections. The observed and derived properties of BX610 are shown in Table \ref{tab:tab01}.

\subsection{ALMA} \label{sec:alma}
    ALMA Band 4 observations of BX610 were carried out in July 2021 as part of Cycle 7 (Project ID: 2019.1.01362.S, PI: R. Herrera-Camus). The details of the data calibration and reduction were presented in \cite{Genzel+23} and \cite{Lee+24}. In summary, we centered two spectral windows (SPWs) set in the Frequency-Division Mode to detect the redshifted \co{4}{3} and \ci{1}{0} transitions with channel widths of \val{3.906}{MHz}, and the two remaining SPWs were set in the Time-Division Mode to detect the rest-frame \val{630}{{\mu}m} dust continuum. The quasar J2253+1608 was used as the flux and bandpass calibrator, and the quasar J2350+1106 was used as the phase calibrator. The ALMA array configurations were chosen to achieve a physical resolution comparable to the Toomre length at this redshift. This resulted in observations with projected baselines of 15 m - 3.7 km that produced an angular resolution of \(\approx\,0.2\arcsec\), which at the redshift of BX610 corresponds to \appval{1.6}{kpc}. We integrated for a total of \val{13.6}{hr}, which resulted in a total observing time of \simval{20}{hr} when including the overheads.

    The data were processed using the Common Astronomy Software Applications package \citep[\texttt{CASA};][]{McMullin+07} version 6.1.1.15. For visibility calibration, we used the pipeline script provided by the ALMA Regional Center staff. Continuum and cube images were produced by the \texttt{TCLEAN} task and deconvolved down to \sig{2} noise level, initially measured from the dirty map. We finally produced \co{4}{3} and \ci{1}{0} line cubes and a \val{630}{\mu m} dust continuum map with a matched, circular beam of \(0.18\arcsec \times 0.18\arcsec\) using Briggs weighting.

    To obtain the \co{4}{3} and \ci{1}{0} integrated intensity maps, we used the \textsc{PYTHON} package \texttt{BETTERMOMENTS}\footnote{\url{https://github.com/richteague/bettermoments}} \citep{Teague+18}, applying a mask based on a \sig{2} clipping of the \(\rm 2 \times FWHM\) convolved spatial data. The resulting integrated intensity maps of the \co{4}{3} and \ci{1}{0} transitions and the dust continuum are shown in Figure \ref{fig:fig01}. Using the 2D-Gaussian fitting task \texttt{IMFIT}, the integrated flux of the dust continuum emission was found to be \(0.18 \pm 0.01 ~ {\rm mJy}\). As discussed by \cite{Lee+24}, this flux value is consistent with previous ALMA Band 4 observations (\#2013.1.00059S, \#2017.1.00856.S, \#2017.1.01045.S) within the errors, but a factor of \(\sim\,2\) lower than the flux reported by \cite{Brisbin+19} based on PdBI observations.

\subsection{Very Large Telescope}
    We include in our study VLT/SINFONI Adaptive Optics (AO)-assisted \(\rm H{\alpha}\) observations from the SINS/zC-SINF survey \citep[e.g.,][]{ForsterSchreiber+09, ForsterSchreiber+14, ForsterSchreiber+18}. We refer to \cite{ForsterSchreiber+18} for a detailed description of the observation and further data reduction. In summary, the galaxy was observed in the \textit{K}-band for an on-source time of 8.3 hr in LGS-SE AO mode. The achieved angular resolution was \(0.24\arcsec\) (a physical resolution of \appval{2}{kpc}). To obtain the \(\rm H{\alpha}\) flux map, we followed the line fitting described in \cite{ForsterSchreiber+18}. The right-most panel of Figure \ref{fig:fig01} shows the integrated \(\rm H{\alpha}\) map. The emission peaks off-center, on a region of \(\sim\)kiloparsec size in the southwest, and is faint in the central region of BX610 as expected from dust obscuration.

\subsection{Hubble Space Telescope}
    We include HST imaging of the rest-frame optical light based on WFC3/IR observations using the F160W and F110W filters. A detailed description of the observations is provided by \cite{Tacchella+15}. We also use the stellar mass and visual extinction \((A_{\rm V})\) maps derived based on optical to NIR broadband spectral energy distributions as described in the same study.


\section{Methods} \label{sec:sec03}
\subsection{Stacking the [CI] Spectra}
    Since one of our main goals is to analyze radial trends in the molecular gas content and excitation properties in BX610, we need to maximize the detectability of the \co{4}{3} and \ci{1}{0} emission lines across the entire galaxy. However, the signal-to-noise ratio (S/N) of the \ci{1}{0} emission line decreases more rapidly than that of the \co{4}{3} line towards the outer regions. To enhance the detection of the \ci{1}{0} line, we employ a spectral stacking method as described by, for example, \cite{Schruba+11} and \cite{Villanueva+21}.

    The stacking procedure is as follows. First, we determine the central velocity of each spaxel in the \co{4}{3} cube. Next, each \ci{1}{0} spaxel is shifted according to the \co{4}{3} velocity field, to align the spectrum of each line of sight to zero velocity. The spectra are then averaged into radial bins defined by the galactocentric radius, with a bin width of \(0.09\arcsec\)—half the beamsize of the data. This approach ensures that the radial bins do not oversample the radial profiles. To extract the line intensity, we fit a single-Gaussian profile to the averaged-stacked spectra. The uncertainties for each radial bin are calculated using the following equation,
    \begin{equation}
        u = \sigma \sqrt{N} \Delta v,
    \end{equation}
    where \(\sigma\) is the RMS of the emission-free portion of the stacked spectrum, \(N\) is the number of channels included in the mask, and \(\Delta v\) is the channel width in \(\rm km~s^{-1}\). To ensure reliable detection, we only consider \ci{1}{0} averaged-stacked spectra that exceed the \(5 \sigma\) level. Without stacking, the \ci{1}{0} line is initially detected up to 0.7 times one effective stellar radius. By stacking the data, on the other hand, we extend the reliable detection range beyond one stellar effective radius.

\subsection{Molecular Gas Mass}
    Molecular gas masses are computed from the \co{4}{3} and \ci{1}{0} line luminosities, which are defined by \cite{Solomon+05} as,
    \begin{equation}
        L'_{\rm line} ~ {\rm \Big[ K ~ km ~ s^{-1} ~ pc^{2} \Big]} = 3.25 \times 10^{7} S_{\rm line} ~ {\Delta}{v} ~ \nu_{\rm obs}^{-2} ~ (1 + z)^{-3} ~ D_{\rm L}^{2},
    \end{equation}
    where \(S_{\rm line} {\Delta}{v}\) is the measured velocity-integrated line flux in \(\rm Jy~km~s^{-1}\), \(\nu_{\rm obs}\) is the observed line frequency in \(\rm GHz\), \(z\) is the redshift of the source, and \(D_{\rm L}\) is the luminosity distance in \(\rm Mpc\). 
    
    The total cold molecular gas mass, \(M_{\rm mol}\), is then derived from the \co{4}{3} line luminosity as,
    \begin{equation}
        M_{\rm mol} ~ {\rm \big[M_{\odot}\big]} = \alpha_{\rm CO} ~ R_{14} ~ L'_{\rm CO~(4-3)},
    \end{equation}
    where \(\alpha_{\rm CO}\) is the luminosity-to-molecular-gas-mass conversion factor, including a 36\% contribution from helium, and \(R_{14} = L'_{\rm CO~(1-0)} / L'_{\rm CO~(4-3)}\) is the line luminosity ratio between the \(J = 4 \to 3\) and \(J = 1 \to 0\) transitions.

    Several studies \citep[e.g.,][]{Leroy+11, Genzel+12, Narayanan+12, Bolatto+13, Sandstrom+13} have shown that the \(\alpha_{\rm CO}\) conversion factor can result in a wide range of values depending on the galaxy properties, including the density, temperature, and metallicity of the molecular clouds. While for local (U)LIRGs and starburst galaxies a value of \(\alpha_{\rm CO} = 0.8 ~ {\rm M_{\odot}~(K~km~s^{-1}~pc^{2})^{-1}}\) is typically assumed \citep[e.g.,][]{Downes+98}, high redshift star-forming galaxies at the massive end of the main-sequence tend to have \(\alpha_{\rm CO}\) values comparable to the standard Milky Way value of \(\alpha_{\rm MW} = 4.36 ~ {\rm M_{\odot}~(K~km~s^{-1}~pc^{2})^{-1}}\) \citep{Bolatto+13, Tacconi+20}.

    To derive the \(\alpha_{\rm CO}\) value for BX610, we follow the calibration presented in \cite{Genzel+15} \citep[also adopted in][]{Tacconi+18}, which combines the recipes given by \cite{Genzel+12} and \cite{Bolatto+13} through the geometric mean of both empirical functions. The \(\alpha_{\rm CO}\) value based on the metallicity of the gas is given by,
    \begin{align}
        \alpha_{\rm CO} & = \alpha_{\rm MW} \sqrt{10^{-1.27 \times \left(12 + \log{\rm (O/H)} - 8.67\right)}} \nonumber\\
                        & \qquad \qquad \quad \times \sqrt{0.67 \exp{\left[0.36 \times 10^{- \left(12 + \log{\rm (O/H)} - 8.67\right)}\right]}}
    \end{align}
    where \(\rm 12 + \log{(O/H)}\) is the metallicity on the \cite{Pettini+04} calibration scale, denoted as,
    \begin{equation}
        {\rm 12 + \log{(O/H)}}_{\rm PP04} = a - 0.087 \left[\log{(M_{\star}) - b}\right]^{2},
    \end{equation}
    with \(a = 8.74 \pm 0.06\) and \(b = (10.4 \pm 0.05) + (4.46 \pm 0.3) \times \log{(1 + z)} - (1.78 \pm 0.4) \times [\log{(1 + z)}]^{2}\) \citep[][and references within]{Genzel+15}. Based on the total stellar mass of BX610, we find a metallicity value of \(\rm 12 + \log{(O/H)} = 8.65 \pm 0.07\), which yields an \(\alpha_{\rm CO} = 4.51 \pm 0.60 ~ {\rm M_{\odot}~(K~km~s^{-1}~pc^{2})^{-1}}\).

    The \(R_{14}\) line luminosity ratio converts the observed \co{4}{3} line luminosity into the \co{1}{0} line luminosity. \cite{Bolatto+15} conducted \co{1}{0} observations on BX610 using the JVLA and detect a total line luminosity of \(L'_{\rm CO~(1-0)} = (2.6 \pm 0.2) \times 10^{10} ~ {\rm K~km~s^{-1}~pc^{2}}\), which yields a line luminosity ratio of \(R_{14} = 1.1 \pm 0.1\). Under these assumptions and values, we find a molecular gas mass of \(M_{\rm mol} = (1.2 \pm 0.2) \times 10^{11} ~ {\rm M_{\odot}}\).
 
    Alternatively, \(M_{\rm mol}\) can be derived from the \ci{1}{0} emission line \citep[e.g.,][]{Weiss+05, Walter+11, Alaghband-Zadeh+13, Bothwell+17, Popping+17, Valentino+18}, based on an estimated atomic mass of carbon, \(M_{\rm [CI]}\), and the assumption of a given neutral atomic carbon abundance \(X{\rm [CI]} / X{\rm [H_{2}]}\). The atomic carbon mass is estimated from the \ci{1}{0} line luminosity emission following the formula presented by \cite{Weiss+05},
    \begin{equation}
        M_{\rm [CI]} ~ {\rm \big[M_{\odot}\big]} = 5.706 \times 10^{-4} ~ Q(T_{\rm ex}) ~ \frac{\exp{(T_{1} / T_{\rm ex})}}{3} ~ L'\:\!\!_{\rm [CI]~(1-0)},
    \end{equation}
    where \(Q(T_{\rm ex}) = 1 + 3 \exp{(-T_{1} / T_{\rm ex})} + 5 \exp{(T_{2} / T_{\rm ex})}\) is the partition function, \(T_{\rm ex}\) is the excitation temperature, and \(T_{1} = 23.6 ~ {\rm K}\) and \(T_{2} = 62.5 ~ {\rm K}\) are the transition energy levels above ground state. \cite{Brisbin+19} recently measured \(T_{\rm ex}\) for BX610 based on the \(L'\:\!\!_{\rm [CI]~(2-1)} ~ \big/ ~ L'\:\!\!_{\rm [CI]~(1-0)}\) line luminosity ratio and obtained a value of \(T_{\rm ex} = 31.8 \pm 6.9 ~ {\rm K}\), which is comparable to the typical dust temperature of \(T_{\rm dust} \sim 32 ~ {\rm K}\) found in massive \((M_{\star} \sim 10^{11}-10^{11.5} ~ {\rm M_{\odot}})\), main-sequence galaxies at \(z \sim 2\) \citep{Magnelli+14, Genzel+15, Schreiber+18}. We then estimate an atomic carbon mass value of \(M_{\rm [CI]} = (8.3 \pm 0.9) \times 10^{6} ~ {\rm M_{\odot}}\). 

    To convert the carbon mass into a molecular gas mass, we need to assume a neutral atomic carbon abundance, \(X{\rm [CI]} / X{\rm [H_{2}]} = M_{\rm [CI]} / 6 M_{\rm H_{2}}\). For instance, \cite{Valentino+18} compare the molecular hydrogen masses derived from CO and dust in a sample of main-sequence galaxies at \(z \sim 1.2\) and find abundances that range between \(1-13 \times 10^{-5}\), with a weighted mean of \(1.5 \times 10^{-5}\). Assuming this mean abundance, we measure a molecular hydrogen mass of \(M_{\rm H_{2}} = (9.5 \pm 0.7) \times 10^{10} ~ {\rm M_{\odot}}\), which results in a molecular gas mass of \(M_{\rm mol} = (1.3 \pm 0.1) \times 10^{11} ~ {\rm M_{\odot}}\) (i.e., consistent with our CO-based measurement).

    These CO- and [CI]-based molecular gas masses agree with the value reported previously by \cite{Bolatto+15}, who estimated a molecular gas mass of \(M_{\rm mol} = (1.1 \pm 0.1) \times 10^{11} ~ {\rm M_{\odot}}\). However, these are a factor of \(\sim\,2\) lower than the value found by \cite{Brisbin+19}, who reported a molecular gas mass of \(M_{\rm mol} = (2.1 \pm 0.3) \times 10^{11} ~ {\rm M_{\odot}}\). This could be explained because \cite{Brisbin+19}, as described in Section \ref{sec:alma}, measured a PdBI dust continuum that is a factor of \(\sim\,2\) larger than the value reported in this work and in \cite{Lee+24}.

\subsection{Star Formation Rate}
    We measure the SFR from the \(\rm H{\alpha}\) line using the calibration by \cite{Kennicutt+98a}, modified for a \cite{Chabrier+03} IMF, as follows,
    \begin{equation}
        {\rm SFR_{H{\alpha}} ~ \Big[ M_{\odot} ~ yr^{-1} \Big]} = 4.7 \times 10^{-42} ~ L_{\rm H{\alpha}, \,corr},
    \end{equation}
    where \( L_{\rm H{\alpha}, \,corr}\) is the \(\rm H{\alpha}\) luminosity in \(\rm erg~s^{-1}\) corrected for attenuation \(A_{\rm H\alpha}\), i.e., \(L_{\rm H{\alpha}, \,corr} = L_{\rm H{\alpha}, \,obs} \times 10^{0.4 A_{\rm H{\alpha}}}\). For the attenuation \(A_{\rm H\alpha}\), we use the extinction radial profiles measured by \cite{Tacchella+18} based on HST observations of the F438W \((B)\) \(-\) F814W \((I)\) index color. We assume a \(E(B-V)_{\rm star} / E(B-V)_{\rm gas}\) ratio of 0.44 \citep[from][]{Calzetti+00}. Under these assumptions, we obtain a SFR of \(140 ~ {\rm M_{\odot}~yr^{-1}}\).
    \begin{figure*}[!ht]
        \centering
        \includegraphics[scale=0.99]{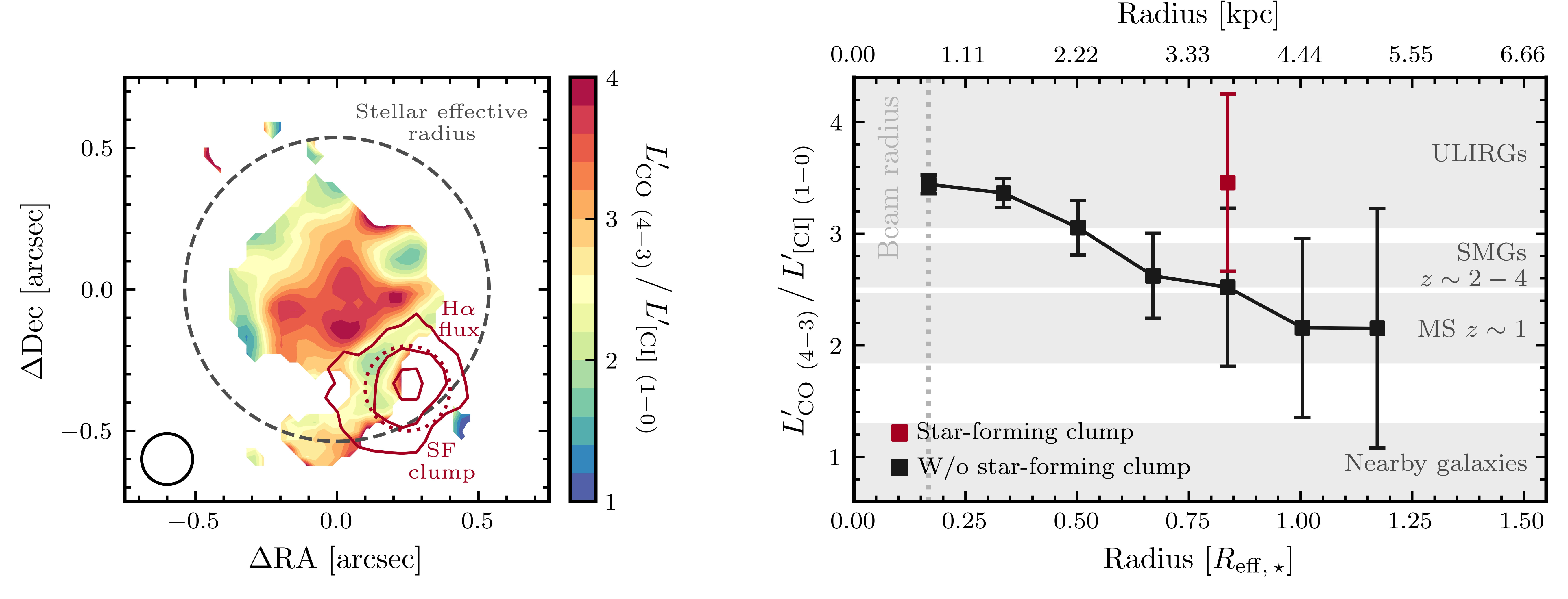}
        \caption{\textit{Left:} \(L'_{\rm CO~(4-3)} ~ \big/ ~ L'\:\!\!_{\rm [CI]~(1-0)}\) line luminosity ratio map. The galaxy's stellar effective radius is shown as the dashed gray circle. The red solid contours correspond to \(\rm H{\alpha}\) fractional flux levels relative to the maximum of 0.5, 0.75, and 0.95, while the red dotted contour denotes the position of the star-forming clump. The filled circle at the bottom left corner indicates the beam size. \textit{Right:} Radial profile of the \(L'_{\rm CO~(4-3)} ~ \big/ ~ L'\:\!\!_{\rm [CI]~(1-0)}\) line luminosity ratio. The radial distance from the center is normalized by the galaxy's stellar effective radius. The black squares with error bars show the mean value of the line luminosity ratio within each beam-sized bin, without considering the emission of the clump. The red square with the error bar shows the mean value of the line luminosity ratio within the clump. The gray-shaded areas are typical line luminosity ratio values, taken from the literature \citep[e.g.,][]{Papadopoulos+12, Valentino+20}. The vertical gray dotted line indicates the beam radius. We observe that the mean value in the central region is comparable to that of local ULIRGs, as well as the mean value in the star-forming clump, while the mean value in the outer disk is comparable to that of the main-sequence galaxies at \(z \sim 1\).}
        \label{fig:fig02}
    \end{figure*}
    Additionally, in dusty environments such as the center of galaxies, the luminosity of young massive stars is frequently absorbed and then re-emitted in the IR, with an emission peak near \val{100}{{\mu}m} \citep[e.g.,][]{Lutz+16}. Therefore, the infrared emission can be a proxy for estimating the dust-obscured star formation. As can be seen from Figure \ref{fig:fig01}, the molecular gas content as traced by the \co{4}{3} and \ci{1}{0} transitions peaks at the center of the galaxy; however, the \(\rm H{\alpha}\) line in this region is very faint. This is the result of strong dust obscuration in the center, where a correction based on the attenuation \(A_{\rm H{\alpha}}\) measured from HST imaging is likely to underestimate the total star formation activity significantly. For this reason, in the central region of BX610, we use the ALMA dust continuum to measure the obscured star formation. First, we scale the ALMA Band 4 dust continuum flux to calculate the rest-frame \val{70}{{\mu}m} flux assuming a modified black-body \citep[MBB,][]{Casey+12} with a spectral emissivity index \(\beta = 2\) and a dust temperature \(T_{\rm dust} = 30 ~ {\rm K}\), as determined by \cite{Brisbin+19}. The resulting flux is \(S_{\rm cont, \,70 ~ {\mu}m} = 13.2 ~ {\rm mJy}\). Finally, we use the calibration by \cite{Calzetti+10} to measure the SFR based on the \val{70}{{\mu}m} continuum emission, which is given by,
    \begin{equation}
        {\rm SFR_{70} ~ \Big[ M_{\odot} ~ yr^{-1} \Big]} = 5.9 \times 10^{-44} ~ L_{70},
    \end{equation}
    where \(L_{70}\) is the luminosity at \val{70}{{\mu}m}. Under these assumptions, we obtain a SFR of \(114 ~ {\rm M_{\odot}~yr^{-1}}\). Nevertheless, the dust continuum-based method has its own limitations. If we assume a dust temperature of \(T_{\rm dust} = 40 ~ {\rm K}\), the resulting SFR increases by a factor of 3.8. On the other hand, assuming a spectral emissivity index of \(\beta = 1.5\), the SFR decreases by a factor of 3.0. Despite all these limitations, if we compare the obscured and unobscured SFR as traced by the dust continuum and uncorrected \(\rm H{\alpha}\) emission, respectively, we measure an obscured fraction of star formation of \(0.84\). This is consistent with what has been observed in other massive star-forming galaxies at \(z \sim 2\) \citep[e.g.,][]{Bethermin+15}.


\section{Results} \label{sec:sec04}
    \subsection{Molecular Gas Excitation}
    With the development of wideband receivers that can simultaneously detect the \co{4}{3} and \ci{1}{0} transitions in galaxies within certain redshift ranges (such as BX610 at \(z \approx 2\)), the \(L'_{\rm CO~(4-3)} ~ \big/ ~ L'\:\!\!_{\rm [CI]~(1-0)}\) line luminosity ratio has become an important tracer of both molecular gas chemistry and excitation \citep[e.g.,][]{Alaghband-Zadeh+13, Bisbas+15, Bothwell+17, Valentino+18, Valentino+20, Michiyama+21}. However, considering that the excitation temperature of the \ci{1}{0} emission line is only about 1.5 times higher than that of the \co{2}{1} emission line, and the critical density of the \ci{1}{0} line is approximately 4.4 times lower than that of the \co{1}{0} transition \citep[e.g.,][]{Carilli+13}, we expect molecular gas excitation to be the dominant effect traced by the \(L'_{\rm CO~(4-3)} ~ \big/ ~ L'\:\!\!_{\rm [CI]~(1-0)}\) line luminosity ratio.
    \begin{figure*}[!ht]
        \centering
        \includegraphics{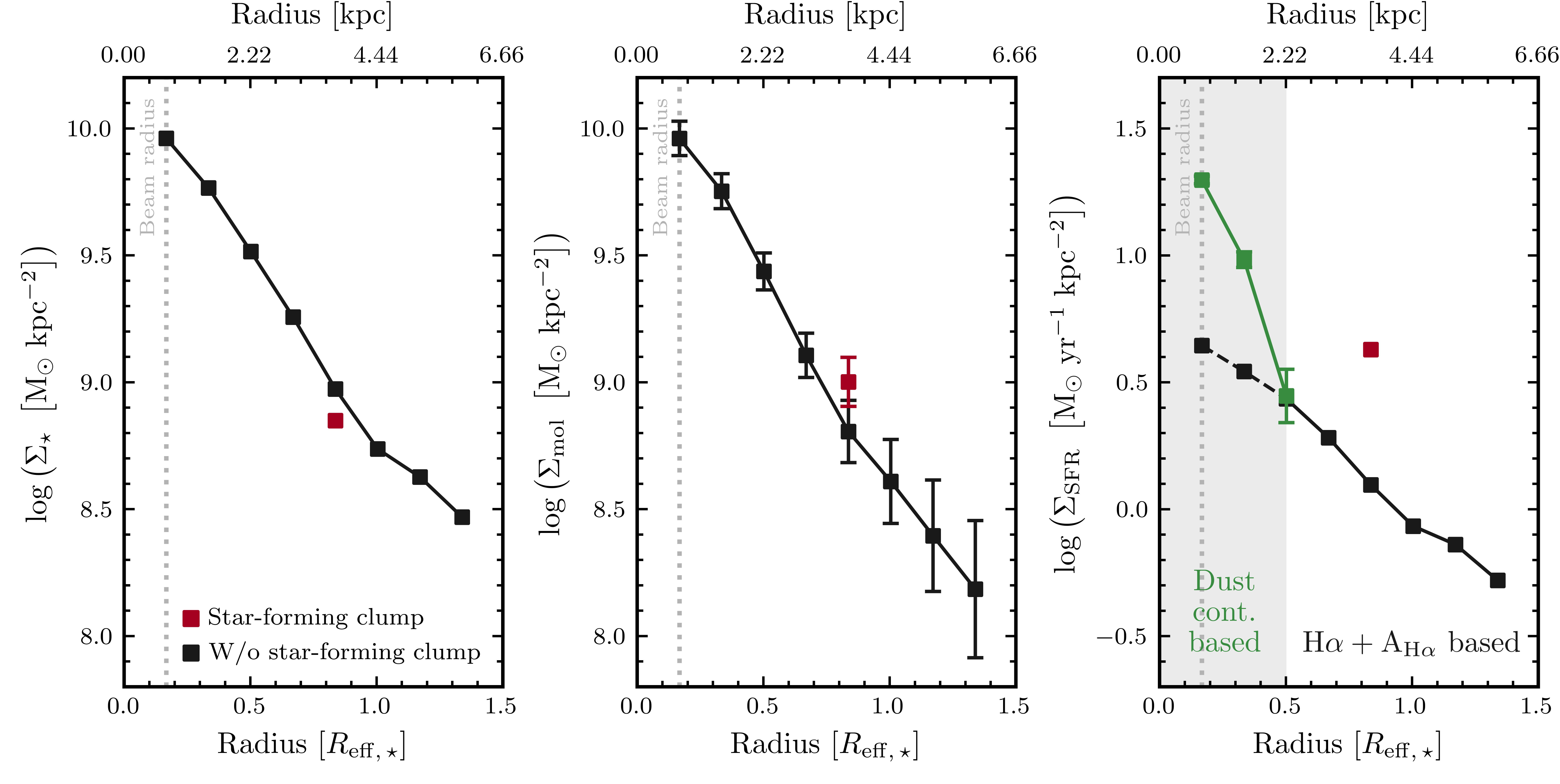}
        \caption{Radial profiles of the stellar mass (left), molecular gas mass (middle), and \(\rm SFR\) (right) surface densities of BX610, plotted with the same logarithmic dynamic range to facilitate comparison. Colors indicate the same as in Figure \ref{fig:fig02}. In the right panel, green squares above the gray-shaded area are derived from the rest-frame \val{630}{\mu m} dust continuum emission, while black squares outside the area are derived from \(\rm H{\alpha}\) corrected by dust attenuation profile of \cite{Tacchella+18}. We also indicate for comparison the \(\rm \Sigma_{SFR}(H{\alpha})\) data in the central region as black-filled squares, but with a dashed line. The vertical gray dotted line indicates the beam radius. We observe in the central region of the right panel that the \(\rm SFR_{H{\alpha}}\) surface density is clearly underestimated in comparison with the \(\rm SFR_{dust}\) surface density.}
        \label{fig:fig03}
    \end{figure*}
    To date, most studies have focused on global measurements of the \(L'_{\rm CO~(4-3)} ~ \big/ ~ L'\:\!\!_{\rm [CI]~(1-0)}\) line luminosity ratio at high redshift. For example, in main-sequence star-forming galaxies at \(z \sim 1\), \cite{Valentino+20} find an average value of \(L'_{\rm CO~(4-3)} ~ \big/ ~ L'\:\!\!_{\rm [CI]~(1-0)} = 2.13 \pm 0.31\). In the case of BX610, however, we now have the opportunity to explore excitation conditions on kiloparsec scales, extending out to one stellar effective radius thanks to our stacking technique. The left panel of Figure \ref{fig:fig02} presents the \(L'_{\rm CO~(4-3)} ~ \big/ ~ L'\:\!\!_{\rm [CI]~(1-0)}\) line luminosity ratio map, normalized to the average value found in \(z \sim 1\) main-sequence galaxies \citep{Valentino+20}. We note that the molecular gas in the central regions of BX610 is significantly more excited than in the outskirts, with a line ratio nearly twice the typical global value found in \(z \sim 1\) main-sequence galaxies. Interestingly, the spiral arm extending southward, next to the giant star-forming clump detected in \(\rm H\alpha\) emission, exhibits a line ratio comparable to that of \(z \sim 1\) main-sequence galaxies.

    The right panel of Figure \ref{fig:fig02} displays the radial profile of the \(L'_{\rm CO~(4-3)} ~ \big/ ~ L'\:\!\!_{\rm [CI]~(1-0)}\) line luminosity ratio, with the radius normalized to the effective stellar radius in the bottom x-axis. At each radius, the profile is calculated by taking the ratio of the summed \co{4}{3} line luminosities to the stacked \ci{1}{0} line luminosities. The emission from the star-forming clump is masked (based on the area marked by the red dashed aperture on the left panel) and shown separately (in red). In the central region \((\lesssim 0.5 ~ {R_{\rm eff, \,\star}})\) of BX610, the line luminosity ratio is comparable to that found in local (U)LIRGs \citep[e.g.,][]{Papadopoulos+12}. Moving outwards, at around \(r \sim 0.5 ~ {R_{\rm eff, \,\star}}\), the ratio decreases to levels similar to those observed in SMGs at \(z \sim 2-4\) \citep{Valentino+20}. Finally, in the range of \(r \sim [0.75 - 1.25] ~ {R_{\rm eff, \,\star}}\), the line luminosity ratio is comparable to the typical values found in main-sequence star-forming galaxies at \(z \sim 1\) \citep{Valentino+20}. The star-forming clump, located at \(r \sim 0.8 ~ {R_{\rm eff, \,\star}}\), exhibits a line luminosity ratio similar to that in the center of BX610, indicating elevated molecular gas excitation due to intense star formation. Notably, the global line luminosity ratio in BX610, \(L'_{\rm CO~(4-3)} ~ \big/ ~ L'\:\!\!_{\rm [CI]~(1-0)} = 2.7 \pm 0.7\), is consistent with that observed in \(z \sim 1\) main-sequence galaxies \citep{Valentino+20}. This highlights the importance of spatially resolved observations in revealing the full range of molecular excitation conditions within galaxies, especially in regions of elevated star formation activity, such as central areas or extra-nuclear star-forming clumps.
    
    \subsection{Stellar, Molecular, and Star Formation Rate Radial Profiles}
    \begin{figure*}
        \centering
        \includegraphics{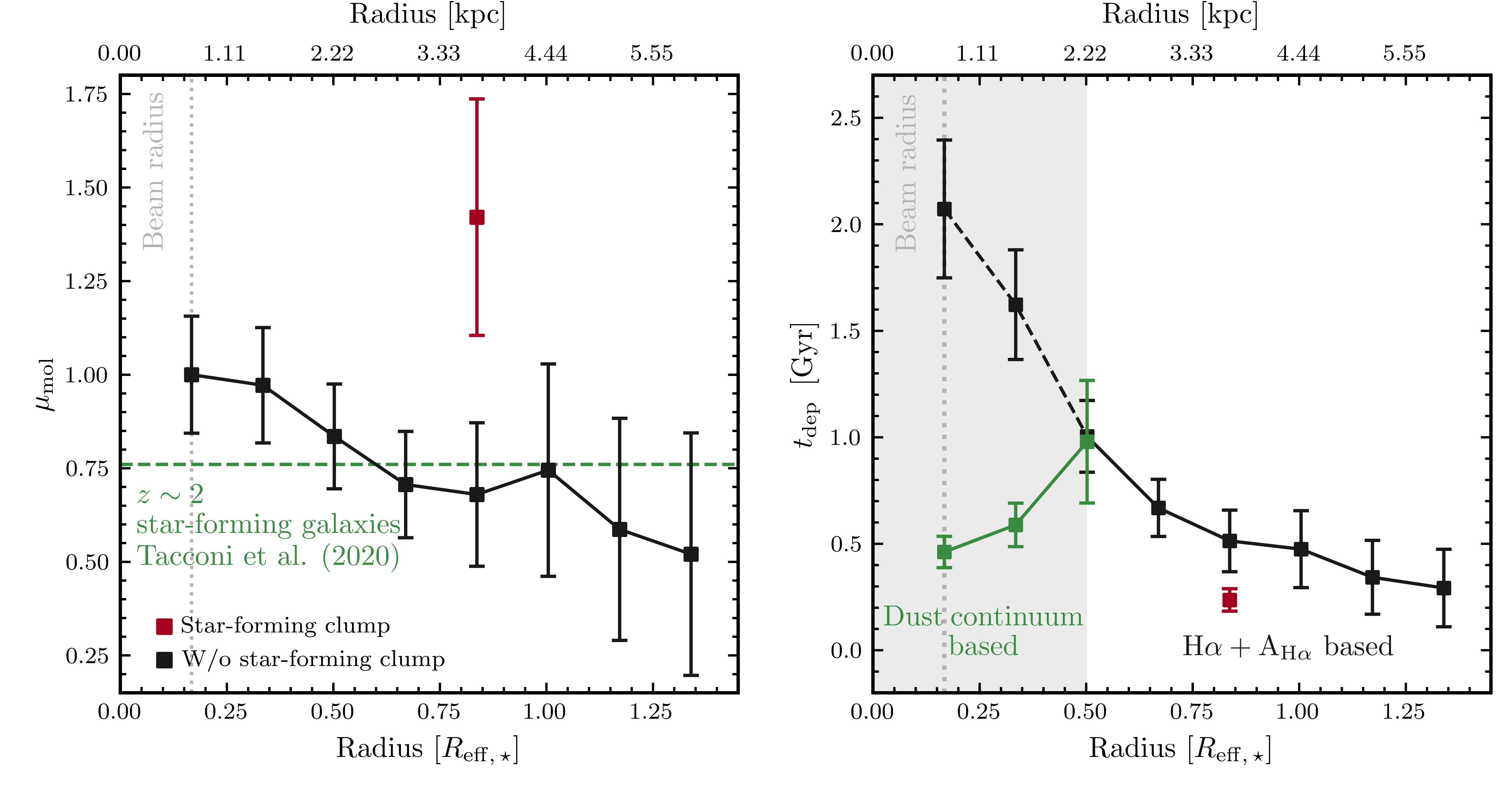}
        \caption{Radial profiles of the molecular gas fraction (left) and the molecular gas depletion time (right) in BX610. Colors indicate the same as in Figure \ref{fig:fig03}. The vertical gray dotted line indicates the beam radius. In the left panel, the green dashed line indicates the average molecular gas fraction measured by \cite{Tacconi+20} for main-sequence galaxies at \(z \sim 2\).}
        \label{fig:fig04}
    \end{figure*}
    Figure \ref{fig:fig03} presents, from left to right, the radial profiles of stellar mass \((\Sigma_{\star})\), molecular gas mass \((\Sigma_{\rm mol})\), and SFR \((\Sigma_{\rm SFR})\) surface densities in BX610. Similar to Figure \ref{fig:fig02}, we mask the emission from the star-forming clump in all three panels and represent it as a single red point. The middle panel shows the molecular gas radial profile based on the CO emission. We choose the CO-based over the [CI]-based profile because CO emission extends farther across the disk, enabling the study of molecular gas surface density up to a radius of \simval{1.5}{\textit{R}_{eff, \,\star}}. In the right panel, we present the radial profile of the SFR derived from extinction-corrected \(\rm H{\alpha}\) emission (black squares) and from dust continuum emission (green squares). In the central region \((\lesssim 0.5 ~ {R_{\rm eff, \,\star}})\), where ALMA detects dust continuum and \(\rm H{\alpha}\) emission is faint, the dust continuum-based SFR is approximately five times higher than the extinction-corrected \(\rm H{\alpha}\)-based SFR. This discrepancy is likely due to the optical extinction correction from \cite{Tacchella+18} being insufficient to fully account for the significant dust obscuration in the central region of BX610. However, it is important to note that the infrared-based SFR here is derived from dust continuum observations at a wavelength away from the peak of the infrared SED, introducing significant uncertainties. To obtain a more accurate measurement of the total SFR in the central region of BX610, ALMA dust continuum observations closer to the SED peak, such as those in Bands 8 or 9 (rest-frame \simval{150}{{\mu}m}), are required. 

    Additionally, we observe that both the star-forming clump and the disk have similar stellar mass surface densities values at \(r \sim 0.8 ~ {R_{\rm eff, \,\star}}\). However, the clump shows slightly higher average values of \(\Sigma_{\rm mol}\) and significantly higher values of \(\Sigma_{\rm SFR}\) at this galactocentric distance than the latter. This suggests that the elevated star formation activity in the clump is a result of the greater concentration of molecular gas compared to other regions of the disk. The similarity in \(\Sigma_{\star}\) values between the disk and the clump at \(r \sim 0.8 ~ {R_{\rm eff, \,\star}}\) suggests that the star formation activity is so recent that it has not yet led to a significant increase in the stellar mass of the clump region.

    The left panel of Figure \ref{fig:fig04} shows the radial profile of the molecular gas fraction, \(\mu_{\rm mol} = M_{\rm mol} / M_{\star}\), which serves as an indicator of the gas available for star formation relative to the existing stellar content. We observe that in the central region \((\lesssim 0.4 ~ {R_{\rm eff, \,\star}})\) of BX610, the molecular gas fraction is close to unity, and then it decreases to \(\mu_{\rm mol} \approx 0.5\) beyond one stellar effective radius. The global molecular gas fraction in BX610 is comparable to that measured in main-sequence star-forming galaxies at \(z \sim 2\) \citep[\(\mu_{\rm mol} \approx 0.75\);][]{Tacconi+20}, but this is a result of averaging the high molecular gas fraction of \(\mu_{\rm mol} \approx 1\) measured at the central \simval{2}{kpc} of BX610 with the \(\mu_{\rm mol} \approx 0.5\) in the outer parts. The star-forming clump in the southwest has an even higher molecular gas fraction than the central region \((\mu_{\rm mol} \approx 1.4)\), which is consistent with the elevated star formation activity traced by the \(\rm H{\alpha}\) line. However, it is important to note that in the dusty central region of BX610, the SED-derived stellar mass may be underestimated.

    The right panel of Figure \ref{fig:fig04} shows the molecular gas depletion time, \(t_{\rm dep} = M_{\rm mol} / {\rm SFR}\), which represents the time required to deplete the molecular gas reservoirs at the current rate of star formation. In the inner \simval{0.5}{\textit{R}_{eff, \,\star}} (\simval{2}{kpc}) radius, the depletion timescale varies depending on whether we assume the extinction-corrected \(\rm H{\alpha}\)-based (black squares) or the dust continuum-based SFR (green squares). Given the elevated concentration of molecular gas in the center and the strong dust continuum detected with ALMA, it is possible that the latter may be more representative of the total SFR in the nuclear region of BX610. If that is the case, then the depletion time decreases towards the center, reaching values of \(\approx 450 ~ {\rm Myr}\), which is consistent with depletion timescales measured in massive, star-forming galaxies at \(z \sim 2\) \cite[e.g.,][]{Tacconi+20}. The star-forming clump is the region with the lowest depletion time in BX610: the current level of star formation activity should exhaust the molecular gas in only \appval{250}{Myr}. Moreover, the estimates derived on the timescales for gas to be depleted by \(\rm SFR\) could be even shorter, given that ionized gas outflows are also observed at those locations \citep{ForsterSchreiber+14}. 

    Placing these results in the broader context of galaxy evolution at cosmic noon, we infer that BX610 is likely in a phase where the rapid inflow of molecular gas into the central region, as reported by \cite{Genzel+23}, has created a gas-rich nucleus \((\mu_{\rm mol} \approx 1)\), driving a high central star formation rate. This phase is commonly referred to as the wet compaction \citep{Dekel+14, Zolotov+15, Tacchella+16b}. Following this phase, galaxies are expected to cease star formation in their centers before the outskirts, driven by a steep decline in the molecular gas fraction in the central kiloparsec region—a pattern opposite to what is observed in BX610 but consistent with findings in compact star-forming galaxies at \(z \sim 2\) in the process of inside-out quenching \citep[e.g.,][]{Spilker+19}.
    
    \subsection{Kennicutt-Schmidt Law}
    \begin{figure}
        \centering
        \includegraphics{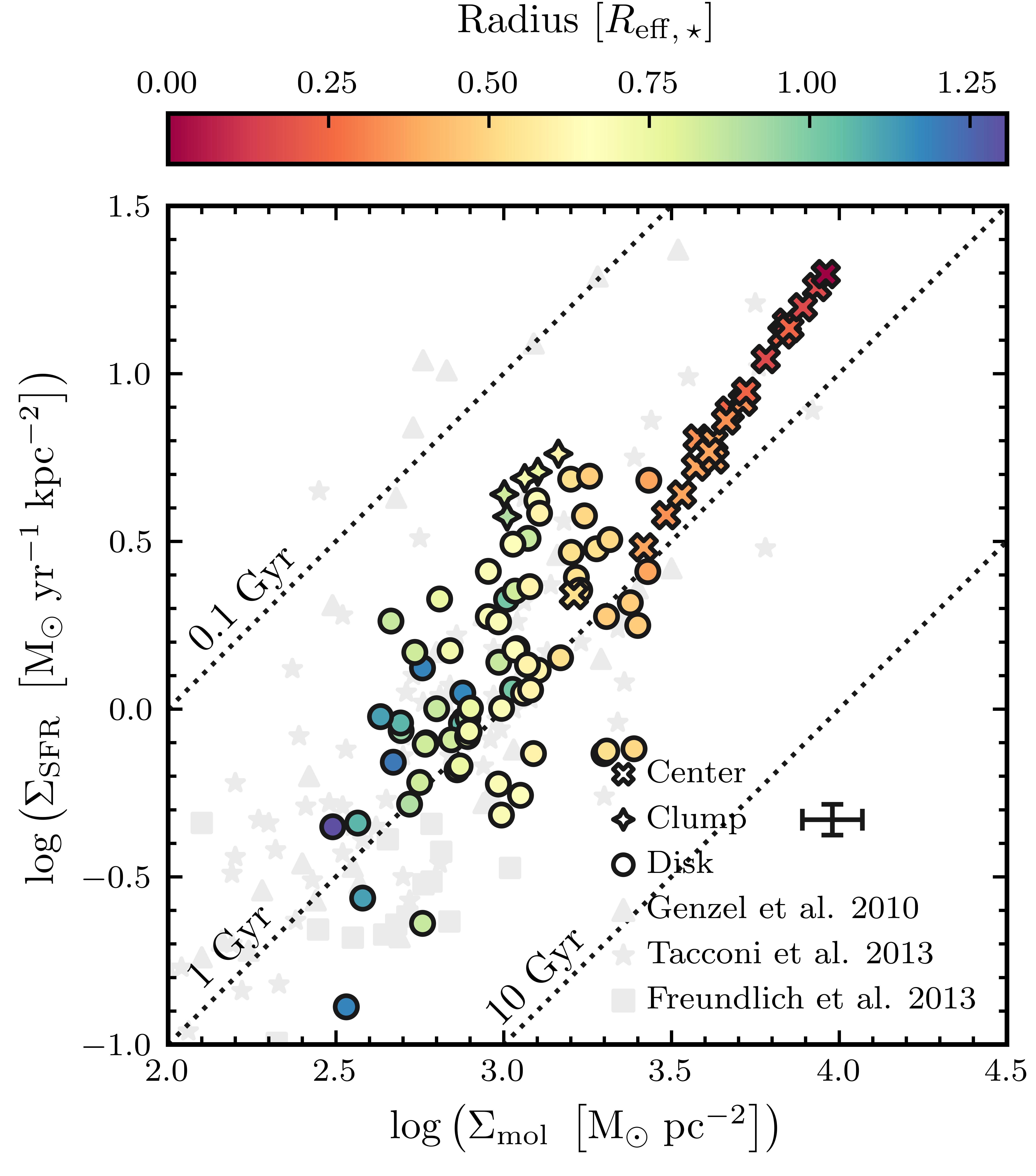}
        \caption{Kennicutt-Schmidt diagram of BX610 for \(0.18\arcsec\) overlapped apertures of the center (crosses), the clump (stars), and the disk (circles). SFR values for disk and clump regions are derived from extinction-corrected \(\rm H{\alpha}\) flux, while for central regions, they are based on dust continuum emission. The color bar indicates the distance to the center as a function of the stellar effective radius. A typical error bar is shown in the bottom right corner. The dotted diagonal lines correspond to constant gas depletion times of \(0.1\), \(1\), and \(\rm 10 ~ Gyr\) from top to bottom. The gray data points from \cite{Genzel+10} and \cite{Tacconi+13} are indicated for comparison with whole galaxies, while the data points from \cite{Freundlich+13} are for spatially resolved main-sequence galaxies at \(z \sim 1\).}
        \label{fig:fig05}
    \end{figure}
    We explore the relationship between molecular gas surface density \((\Sigma_{\rm mol})\) and star formation rate surface density \((\Sigma_{\rm SFR})\), as described by the Kennicutt-Schmidt law \citep{Schmidt+59, Kennicutt+98a}. This relationship is key to understanding how efficiently cold gas is converted into stars across different environments \citep[e.g.,][]{Genzel+10, Freundlich+13, Tacconi+13, Hodge+15, Chen+17, Sun+23, Villanueva+24}. Figure \ref{fig:fig05} shows the resolved KS relation for BX610, with data points color-coded by radial distance and distinguished by shape according to specific regions within the galaxy. The figure is based on measurements from apertures across BX610, each with a diameter of \(0.18\arcsec\) (matching the beam size) and centers separated by \(0.09\arcsec\) to minimize flux loss from overlapping apertures. For the disk and clump regions, SFR values are derived from extinction-corrected \(\rm H{\alpha}\) fluxes, while for the central regions, they are based on the ALMA dust continuum emission. Note that the use of \(\rm H{\alpha}\)-based \(\rm SFR\) for the center regions would result in a higher value of the molecular gas depletion time, as can be seen in the right panel of Figure \ref{fig:fig04}.

    Overall, the \(\sim\)kiloparsec-scale regions in BX610 align with the trend observed in spatially resolved studies of other massive, star-forming galaxies at \(z \sim 1-2\) \citep{Tacconi+13, Freundlich+13, Genzel+13}. Disk regions (circles) exhibit depletion timescales that scatter around \simval{1}{Gyr}, with variations of about a factor of \(\sim\,2\). As regions approach the center of BX610 (crosses), depletion times shorten, reaching close to \simval{450}{Myr}. Notably, the extra-nuclear star-forming clump (stars) is the region in BX610 with the highest star formation efficiency, with a short depletion time of only \simval{250}{Myr}. Applying an ordinary least-squares bisector fit to all the regions in BX610, we find that \(\log{(\Sigma_{\rm SFR})} = (1.27 \pm 0.05) \times \log{(\Sigma_{\rm mol})} - (3.68 \pm 0.17)\).

    Similar to the case of the star-forming galaxy EGS13011166 at \(z \approx 1.5\) \citep{Genzel+13}, the fit will depend on the method by which the correction for dust obscuration is implemented. For EGS13011166, the slope ranges from 0.8 to 1.7, depending on the chosen extinction model. However, for the preferred mixed dust-gas model, \cite{Genzel+13} reported a super-linear slope of \(N = 1.14 \pm 0.1\), which is comparable, within uncertainties and assumptions, to the measured slope in BX610 of \(N = 1.27 \pm 0.1\).

    
\section{Conclusions} \label{sec:sec05}
    We have conducted deep, high-angular-resolution \((0.18\arcsec)\) ALMA Band 4 observations targeting the molecular gas traced by the \co{4}{3} and \ci{1}{0} transitions in BX610, a massive main-sequence galaxy at \(z = 2.21\). These observations, combined with VLT/SINFONI AO-assisted \(\rm H{\alpha}\) emission line data \citep{ForsterSchreiber+18}, have allowed us to characterize the molecular gas properties and star formation activity on \(\sim\)kiloparsec scales, spanning from the central regions to at least \(\sim\,1.5\) (\simval{7}{kpc}) stellar effective radius in this typical star-forming galaxy at cosmic noon.
    
    Our key findings are as follows:
    \begin{enumerate}
        \item By using the \(L'_{\rm CO~(4-3)} / L'\:\!\!_{\rm [CI]~(1-0)}\) line luminosity ratio as a tracer of molecular gas excitation in BX610, we find that the central region and the extra-nuclear, giant star-forming clump exhibit excitation levels comparable to those observed in nearby (U)LIRGs \citep{Papadopoulos+12}. Beyond \simval{0.5}{\textit{R}_{eff, \,\star}}, and on average across the disk, the molecular gas excitation conditions in BX610 resemble those typically found in normal star-forming galaxies at \(z \sim 1\) \citep[e.g., ][]{Valentino+20}.
        \item The molecular gas fraction measured in the inner \simval{2}{kpc} (\simval{0.5}{\textit{R}_{eff, \,\star}}) of BX610 and the giant star-forming clump in the southwest is high \((\mu_{\rm mol} \approx 1)\), resulting in intense star formation activity. Overall, the molecular gas fraction in the disk \((\mu_{\rm mol} \approx 0.5-0.7)\) is comparable to that globally measured in main-sequence, massive star-forming galaxies at \(z \approx 2\) \cite[e.g.,][]{Tacconi+20}. The molecular gas depletion times measured in the central region and in the giant star-forming clump are \simval{450}{Myr} and \simval{250}{Myr}, respectively. 
        \item The rapid transport of molecular gas to the central region of BX610 \citep{Genzel+23}, combined with the high central molecular gas fraction and short depletion timescale, suggests that BX610 may be undergoing a phase similar to the wet compaction scenario described in models by \cite{Dekel+14}, \cite{Zolotov+15}, and \cite{Tacchella+18}. In this phase, the concentration of gas in the galaxy's core drives intense star formation, which eventually leads to the inside-out quenching of star formation.
        \item The \(\sim\)kiloparsec-scale regions in BX610 exhibit significant variations but generally align with the spatially-resolved KS relation observed in other massive, star-forming galaxies at \(z \sim 1-2\) \citep[e.g.,][]{Freundlich+13, Genzel+13, Tacconi+13}. The overall slope and the depletion timescale measured in BX610 are \(N = 1.27 \pm 0.05\) and \(t_{\rm dep}\approx 800 \pm 400 ~ {\rm Myr}\), respectively. 
    \end{enumerate}
    In summary, the deep, spatially-resolved ALMA and VLT/SINFONI observations of BX610 provide one of the most detailed maps of molecular gas and star formation activity in a main-sequence, star-forming galaxy at cosmic noon. However, the absence of reliable measurements of obscured star formation in the central region limits our ability to interpret its physical conditions fully. Future ALMA observations at wavelengths closer to the SED peak (Band 8 or Band 9) or measurements of the Paschen-\(\alpha\) line with JWST/MIRI are essential for addressing this issue and obtaining a more comprehensive understanding of the BX610 central region.

\begin{acknowledgements}
    S.A.-N. and R.H.-C. thank the Max Planck Society for support under the Partner Group project “The Baryon Cycle in Galaxies” between the Max Planck for Extraterrestrial Physics and the Universidad de Concepción. S.A.-N. and R.H.-C. also gratefully acknowledge financial support from ANID BASAL projects FB210003.
    V. V. acknowledges support from the ALMA-ANID Postdoctoral Fellowship under the award ASTRO21-0062.
    This paper makes use of the following ALMA data: ADS/JAO.ALMA\#2019.1.01362.S. ALMA is a partnership of ESO (representing its member states), NSF (USA) and NINS (Japan), together with NRC (Canada), NSTC and ASIAA (Taiwan), and KASI (Republic of Korea), in cooperation with the Republic of Chile. The Joint ALMA Observatory is operated by ESO, AUI/NRAO and NAOJ.
\end{acknowledgements}

\bibliographystyle{aa}
\bibliography{references.bib}

\end{document}